\documentstyle[amssymb,12pt]{article}
\makeatletter
\newcount\@hour\newcount\@minute\chardef\@x10\chardef\@xv60
\def\tcitime{
\def\@time{%
  \@minute\time\@hour\@minute\divide\@hour\@xv
  \ifnum\@hour<\@x 0\fi\the\@hour:%
  \multiply\@hour\@xv\advance\@minute-\@hour
  \ifnum\@minute<\@x 0\fi\the\@minute
  }}%

\@ifundefined{hyperref}{}{}

\@ifundefined{qExtProgCall}{\def\qExtProgCall#1#2#3#4#5#6{\relax}}{}
\def\QCTOpt[#1]#2{%
  \def\QCTOptB{#1}
  \def\QCTOptA{#2}
}
\def\QCTNOpt#1{%
  \def\QCTOptA{#1}
  \let\QCTOptB\empty
}
\def\Qct{%
  \@ifnextchar[{%
    \QCTOpt}{\QCTNOpt}
}
\def\QCBOpt[#1]#2{%
  \def\QCBOptB{#1}
  \def\QCBOptA{#2}
}
\def\QCBNOpt#1{%
  \def\QCBOptA{#1}
  \let\QCBOptB\empty
}
\def\Qcb{%
  \@ifnextchar[{%
    \QCBOpt}{\QCBNOpt}
}
\def\PrepCapArgs{%
  \ifx\QCBOptA\empty
    \ifx\QCTOptA\empty
      {}%
    \else
      \ifx\QCTOptB\empty
        {\QCTOptA}%
      \else
        [\QCTOptB]{\QCTOptA}%
      \fi
    \fi
  \else
    \ifx\QCBOptA\empty
      {}%
    \else
      \ifx\QCBOptB\empty
        {\QCBOptA}%
      \else
        [\QCBOptB]{\QCBOptA}%
      \fi
    \fi
  \fi
}
\newcount\GRAPHICSTYPE
\GRAPHICSTYPE=\z@
\def\GRAPHICSPS#1{%
 \ifcase\GRAPHICSTYPE
   \special{ps: #1}%
 \or
   \special{language "PS", include "#1"}%
 \fi
}%
\def\graffile#1#2#3#4{%
    \leavevmode
    \raise -#4 \BOXTHEFRAME{%
        \hbox to #2{\raise #3\hbox to #2{\null #1\hfil}}}%
}%
\def\draftbox#1#2#3#4{%
 \leavevmode\raise -#4 \hbox{%
  \frame{\rlap{\protect\tiny #1}\hbox to #2%
   {\vrule height#3 width\z@ depth\z@\hfil}%
  }%
 }%
}%
\newcount\draft
\draft=\z@

\newif\ifwasdraft
\wasdraftfalse

\def\GRAPHIC#1#2#3#4#5{%
 \ifnum\draft=\@ne\draftbox{#2}{#3}{#4}{#5}%
  \else\graffile{#1}{#3}{#4}{#5}%
  \fi
 }%
\def\addtoLaTeXparams#1{%
    \edef\LaTeXparams{\LaTeXparams #1}}%
\newif\ifBoxFrame \BoxFramefalse
\newif\ifOverFrame \OverFramefalse
\newif\ifUnderFrame \UnderFramefalse

\def\BOXTHEFRAME#1{%
   \hbox{%
      \ifBoxFrame
         \frame{#1}%
      \else
         {#1}%
      \fi
   }%
}

\def\doFRAMEparams#1{\BoxFramefalse\OverFramefalse\UnderFramefalse\readFRAMEparams#1\end}%
\def\readFRAMEparams#1{%
 \ifx#1\end%
  \let\next=\relax
  \else
  \ifx#1i\dispkind=\z@\fi
  \ifx#1d\dispkind=\@ne\fi
  \ifx#1f\dispkind=\tw@\fi
  \ifx#1t\addtoLaTeXparams{t}\fi
  \ifx#1b\addtoLaTeXparams{b}\fi
  \ifx#1p\addtoLaTeXparams{p}\fi
  \ifx#1h\addtoLaTeXparams{h}\fi
  \ifx#1X\BoxFrametrue\fi
  \ifx#1O\OverFrametrue\fi
  \ifx#1U\UnderFrametrue\fi
  \ifx#1w
    \ifnum\draft=1\wasdrafttrue\else\wasdraftfalse\fi
    \draft=\@ne
  \fi
  \let\next=\readFRAMEparams
  \fi
 \next
 }%
\def\IFRAME#1#2#3#4#5#6{%
      \bgroup
      \let\QCTOptA\empty
      \let\QCTOptB\empty
      \let\QCBOptA\empty
      \let\QCBOptB\empty
      #6%
      \parindent=0pt%
      \leftskip=0pt
      \rightskip=0pt
      \setbox0 = \hbox{\QCBOptA}%
      \@tempdima = #1\relax
      \ifOverFrame
          \typeout{This is not implemented yet}%
          \show\HELP
      \else
         \ifdim\wd0>\@tempdima
            \advance\@tempdima by \@tempdima
            \ifdim\wd0 >\@tempdima
               \textwidth=\@tempdima
               \setbox1 =\vbox{%
                  \noindent\hbox to \@tempdima{\hfill\GRAPHIC{#5}{#4}{#1}{#2}{#3}\hfill}\\%
                  \noindent\hbox to \@tempdima{\parbox[b]{\@tempdima}{\QCBOptA}}%
               }%
               \wd1=\@tempdima
            \else
               \textwidth=\wd0
               \setbox1 =\vbox{%
                 \noindent\hbox to \wd0{\hfill\GRAPHIC{#5}{#4}{#1}{#2}{#3}\hfill}\\%
                 \noindent\hbox{\QCBOptA}%
               }%
               \wd1=\wd0
            \fi
         \else
            \ifdim\wd0>0pt
              \hsize=\@tempdima
              \setbox1 =\vbox{%
                \unskip\GRAPHIC{#5}{#4}{#1}{#2}{0pt}%
                \break
                \unskip\hbox to \@tempdima{\hfill \QCBOptA\hfill}%
              }%
              \wd1=\@tempdima
           \else
              \hsize=\@tempdima
              \setbox1 =\vbox{%
                \unskip\GRAPHIC{#5}{#4}{#1}{#2}{0pt}%
              }%
              \wd1=\@tempdima
           \fi
         \fi
         \@tempdimb=\ht1
         \advance\@tempdimb by \dp1
         \advance\@tempdimb by -#2%
         \advance\@tempdimb by #3%
         \leavevmode
         \raise -\@tempdimb \hbox{\box1}%
      \fi
      \egroup%
}%
\def\DFRAME#1#2#3#4#5{%
 \begin{center}
     \let\QCTOptA\empty
     \let\QCTOptB\empty
     \let\QCBOptA\empty
     \let\QCBOptB\empty
     \ifOverFrame 
        #5\QCTOptA\par
     \fi
     \GRAPHIC{#4}{#3}{#1}{#2}{\z@}
     \ifUnderFrame 
        \nobreak\par #5\QCBOptA
     \fi
 \end{center}%
 }%
\def\FFRAME#1#2#3#4#5#6#7{%
 \begin{figure}[#1]%
  \let\QCTOptA\empty
  \let\QCTOptB\empty
  \let\QCBOptA\empty
  \let\QCBOptB\empty
  \ifOverFrame
    #4
    \ifx\QCTOptA\empty
    \else
      \ifx\QCTOptB\empty
        \caption{\QCTOptA}%
      \else
        \caption[\QCTOptB]{\QCTOptA}%
      \fi
    \fi
    \ifUnderFrame\else
      \label{#5}%
    \fi
  \else
    \UnderFrametrue%
  \fi
  \begin{center}\GRAPHIC{#7}{#6}{#2}{#3}{\z@}\end{center}%
  \ifUnderFrame
    #4
    \ifx\QCBOptA\empty
      \caption{}%
    \else
      \ifx\QCBOptB\empty
        \caption{\QCBOptA}%
      \else
        \caption[\QCBOptB]{\QCBOptA}%
      \fi
    \fi
    \label{#5}%
  \fi
  \end{figure}%
 }%
\newcount\dispkind%

\def\makeactives{
  \catcode`\"=\active
  \catcode`\;=\active
  \catcode`\:=\active
  \catcode`\'=\active
  \catcode`\~=\active
}
\bgroup
   \makeactives
   \gdef\activesoff{%
      \def"{\string"}
      \def;{\string;}
      \def:{\string:}
      \def'{\string'}
      \def~{\string~}
    }
\egroup

\def\FRAME#1#2#3#4#5#6#7#8{%
 \bgroup
 \@ifundefined{bbl@deactivate}{}{\activesoff}
 \ifnum\draft=\@ne
   \wasdrafttrue
 \else
   \wasdraftfalse%
 \fi
 \def\LaTeXparams{}%
 \dispkind=\z@
 \def\LaTeXparams{}%
 \doFRAMEparams{#1}%
 \ifnum\dispkind=\z@\IFRAME{#2}{#3}{#4}{#7}{#8}{#5}\else
  \ifnum\dispkind=\@ne\DFRAME{#2}{#3}{#7}{#8}{#5}\else
   \ifnum\dispkind=\tw@
    \edef\@tempa{\noexpand\FFRAME{\LaTeXparams}}%
    \@tempa{#2}{#3}{#5}{#6}{#7}{#8}%
    \fi
   \fi
  \fi
  \ifwasdraft\draft=1\else\draft=0\fi{}%
  \egroup
 }%
\def\TEXUX#1{"texux"}

\long\def\QQQ#1#2{%
     \long\expandafter\def\csname#1\endcsname{#2}}%
\@ifundefined{QTP}{\def\QTP#1{}}{}
\@ifundefined{QEXCLUDE}{\def\QEXCLUDE#1{}}{}
\@ifundefined{Qlb}{}{}
\@ifundefined{Qlt}{}{}
\long\def\QQA#1#2{}%
\def\QTR#1#2{{\csname#1\endcsname #2}}
\def\EXPAND#1[#2]#3{}%
\def\NOEXPAND#1[#2]#3{}%
\def\LaTeXparent#1{}%
\def\ChildStyles#1{}%
\def\ChildDefaults#1{}%
\def\QTagDef#1#2#3{}%
\@ifundefined{StyleEditBeginDoc}{}{}
\def\QQfnmark#1{\footnotemark}

\@ifundefined{INDEX}{\def\INDEX#1#2{}{}}{}%
\@ifundefined{SUBINDEX}{\def\SUBINDEX#1#2#3{}{}{}}{}%
\@ifundefined{initial}%
   {\def\initial#1{\bigbreak{\raggedright\large\bf #1}\kern 2\p@\penalty3000}}%
   {}%
\@ifundefined{entry}{}{}%
\@ifundefined{primary}{}{}%
\@ifundefined{secondary}{}{}%
\@ifundefined{ZZZ}{}{\makeatletter\input gnuindex.sty\makeatother\makeindex\makeatletter}%
\@ifundefined{abstract}{%
 \def\abstract{%
  \if@twocolumn
   \section*{Abstract (Not appropriate in this style!)}%
   \else \small 
   \begin{center}{\bf Abstract\vspace{-.5em}\vspace{\z@}}\end{center}%
   \quotation 
   \fi
  }%
 }{%
 }%
\@ifundefined{endabstract}{\def\endabstract
  {\if@twocolumn\else\endquotation\fi}}{}%
\@ifundefined{maketitle}{\def\maketitle#1{}}{}%
\@ifundefined{affiliation}{\def\affiliation#1{}}{}%
\@ifundefined{proof}{}{}%
\@ifundefined{endproof}{}{}%
\@ifundefined{newfield}{\def\newfield#1#2{}}{}%
\@ifundefined{chapter}{\def\chapter#1{\par(Chapter head:)#1\par }%
 \newcount\c@chapter}{}%
\@ifundefined{part}{\def\part#1{\par(Part head:)#1\par }}{}%
\@ifundefined{section}{\def\section#1{\par(Section head:)#1\par }}{}%
\@ifundefined{subsection}{\def\subsection#1%
 {\par(Subsection head:)#1\par }}{}%
\@ifundefined{subsubsection}{\def\subsubsection#1%
 {\par(Subsubsection head:)#1\par }}{}%
\@ifundefined{paragraph}{\def\paragraph#1%
 {\par(Subsubsubsection head:)#1\par }}{}%
\@ifundefined{subparagraph}{\def\subparagraph#1%
 {\par(Subsubsubsubsection head:)#1\par }}{}%
\@ifundefined{therefore}{}{}%
\@ifundefined{backepsilon}{}{}%
\@ifundefined{yen}{}{}%
\@ifundefined{registered}{%
   \def\registered{\relax\ifmmode{}\r@gistered
                    \else$\m@th\r@gistered$\fi}%
 \def\r@gistered{^{\ooalign
  {\hfil\raise.07ex\hbox{$\scriptstyle\rm\text{R}$}\hfil\crcr
  \mathhexbox20D}}}}{}%
\@ifundefined{Eth}{}{}%
\@ifundefined{eth}{}{}%
\@ifundefined{Thorn}{}{}%
\@ifundefined{thorn}{}{}%
\@ifundefined{degree}{}{}%
\newdimen\theight
\def\Column{%
 \vadjust{\setbox\z@=\hbox{\scriptsize\quad\quad tcol}%
  \theight=\ht\z@\advance\theight by \dp\z@\advance\theight by \lineskip
  \kern -\theight \vbox to \theight{%
   \rightline{\rlap{\box\z@}}%
   \vss
   }%
  }%
 }%
\def\qed{%
 \ifhmode\unskip\nobreak\fi\ifmmode\ifinner\else\hskip5\p@\fi\fi
 \hbox{\hskip5\p@\vrule width4\p@ height6\p@ depth1.5\p@\hskip\p@}%
 }%
\def\miss{\hbox{\vrule height2\p@ width 2\p@ depth\z@}}%
\def\tcol#1{{\baselineskip=6\p@ \vcenter{#1}} \Column}  %
\def\newfmtname{LaTeX2e}
\def\chkcompat{%
   \if@compatibility
   \else
     \usepackage{latexsym}
   \fi
}

\ifx\fmtname\newfmtname
  \DeclareOldFontCommand{\rm}{\normalfont\rmfamily}{\mathrm}
  \DeclareOldFontCommand{\sf}{\normalfont\sffamily}{\mathsf}
  \DeclareOldFontCommand{\tt}{\normalfont\ttfamily}{\mathtt}
  \DeclareOldFontCommand{\bf}{\normalfont\bfseries}{\mathbf}
  \DeclareOldFontCommand{\it}{\normalfont\itshape}{\mathit}
  \DeclareOldFontCommand{\sl}{\normalfont\slshape}{\@nomath\sl}
  \DeclareOldFontCommand{\sc}{\normalfont\scshape}{\@nomath\sc}
  \chkcompat
\fi

\def\alpha{{\Greekmath 010B}}%
\def\beta{{\Greekmath 010C}}%
\def\gamma{{\Greekmath 010D}}%
\def\delta{{\Greekmath 010E}}%
\def\epsilon{{\Greekmath 010F}}%
\def\zeta{{\Greekmath 0110}}%
\def\eta{{\Greekmath 0111}}%
\def\theta{{\Greekmath 0112}}%
\def\iota{{\Greekmath 0113}}%
\def\kappa{{\Greekmath 0114}}%
\def\lambda{{\Greekmath 0115}}%
\def\mu{{\Greekmath 0116}}%
\def\nu{{\Greekmath 0117}}%
\def\xi{{\Greekmath 0118}}%
\def\pi{{\Greekmath 0119}}%
\def\rho{{\Greekmath 011A}}%
\def\sigma{{\Greekmath 011B}}%
\def\tau{{\Greekmath 011C}}%
\def\upsilon{{\Greekmath 011D}}%
\def\phi{{\Greekmath 011E}}%
\def\chi{{\Greekmath 011F}}%
\def\psi{{\Greekmath 0120}}%
\def\omega{{\Greekmath 0121}}%
\def\varepsilon{{\Greekmath 0122}}%
\def\vartheta{{\Greekmath 0123}}%
\def\varpi{{\Greekmath 0124}}%
\def\varrho{{\Greekmath 0125}}%
\def\varsigma{{\Greekmath 0126}}%
\def\varphi{{\Greekmath 0127}}%

\def\nabla{{\Greekmath 0272}}
\def\FindBoldGroup{%
   {\setbox0=\hbox{$\mathbf{x\global\edef\theboldgroup{\the\mathgroup}}$}}%
}

\def\Greekmath#1#2#3#4{%
    \if@compatibility
        \ifnum\mathgroup=\symbold
           \mathchoice{\mbox{\boldmath$\displaystyle\mathchar"#1#2#3#4$}}%
                      {\mbox{\boldmath$\textstyle\mathchar"#1#2#3#4$}}%
                      {\mbox{\boldmath$\scriptstyle\mathchar"#1#2#3#4$}}%
                      {\mbox{\boldmath$\scriptscriptstyle\mathchar"#1#2#3#4$}}%
        \else
           \mathchar"#1#2#3#4%
        \fi 
    \else 
        \FindBoldGroup
        \ifnum\mathgroup=\theboldgroup 
           \mathchoice{\mbox{\boldmath$\displaystyle\mathchar"#1#2#3#4$}}%
                      {\mbox{\boldmath$\textstyle\mathchar"#1#2#3#4$}}%
                      {\mbox{\boldmath$\scriptstyle\mathchar"#1#2#3#4$}}%
                      {\mbox{\boldmath$\scriptscriptstyle\mathchar"#1#2#3#4$}}%
        \else
           \mathchar"#1#2#3#4%
        \fi     	    
	  \fi}

\newif\ifGreekBold  \GreekBoldfalse
\let\SAVEPBF=\pbf
\def\pbf{\GreekBoldtrue\SAVEPBF}%

\@ifundefined{theorem}{}{}
\@ifundefined{lemma}{}{}
\@ifundefined{corollary}{}{}
\@ifundefined{conjecture}{}{}
\@ifundefined{proposition}{}{}
\@ifundefined{axiom}{}{}
\@ifundefined{remark}{}{}
\@ifundefined{example}{}{}
\@ifundefined{exercise}{}{}
\@ifundefined{definition}{}{}

\@ifundefined{mathletters}{%
  \newcounter{equationnumber}  
  \def\mathletters{%
     \addtocounter{equation}{1}
     \edef\@currentlabel{\theequation}%
     \setcounter{equationnumber}{\c@equation}
     \setcounter{equation}{0}%
     \edef\theequation{\@currentlabel\noexpand\alph{equation}}%
  }
  
}{}

\@ifundefined{BibTeX}{%
    \def\BibTeX{{\rm B\kern-.05em{\sc i\kern-.025em b}\kern-.08em
                 T\kern-.1667em\lower.7ex\hbox{E}\kern-.125emX}}}{}%
\@ifundefined{AmS}%
    {\def\AmS{{\protect\usefont{OMS}{cmsy}{m}{n}%
                A\kern-.1667em\lower.5ex\hbox{M}\kern-.125emS}}}{}%
\@ifundefined{AmSTeX}{}{}%
\ifx\ds@amstex\relax
\makeatother\endinput
\else
   \@ifpackageloaded{amstex}%
      {\message{amstex already loaded}\makeatother\endinput}
      {}
   \@ifpackageloaded{amsgen}%
      {\message{amsgen already loaded}\makeatother\endinput}
      {}
\fi
\def\DN@{\def\next@}%
\def\eat@#1{}%
\let\DOTSI\relax
\def\RIfM@{\relax\ifmmode}%
\def\FN@{\futurelet\next}%
\newcount\intno@
\def\iint{\DOTSI\intno@\tw@\FN@\ints@}%
\def\iiint{\DOTSI\intno@\thr@@\FN@\ints@}%
\def\iiiint{\DOTSI\intno@4 \FN@\ints@}%
\def\idotsint{\DOTSI\intno@\z@\FN@\ints@}%
\def\ints@{\findlimits@\ints@@}%
\newif\iflimtoken@
\newif\iflimits@
\def\findlimits@{\limtoken@true\ifx\next\limits\limits@true
 \else\ifx\next\nolimits\limits@false\else
 \limtoken@false\ifx\ilimits@\nolimits\limits@false\else
 \ifinner\limits@false\else\limits@true\fi\fi\fi\fi}%
\def\multint@{\int\ifnum\intno@=\z@\intdots@                          
 \else\intkern@\fi                                                    
 \ifnum\intno@>\tw@\int\intkern@\fi                                   
 \ifnum\intno@>\thr@@\int\intkern@\fi                                 
 \int}
\def\multintlimits@{\intop\ifnum\intno@=\z@\intdots@\else\intkern@\fi
 \ifnum\intno@>\tw@\intop\intkern@\fi
 \ifnum\intno@>\thr@@\intop\intkern@\fi\intop}%
\def\intic@{%
    \mathchoice{\hskip.5em}{\hskip.4em}{\hskip.4em}{\hskip.4em}}%
\def\negintic@{\mathchoice
 {\hskip-.5em}{\hskip-.4em}{\hskip-.4em}{\hskip-.4em}}%
\def\ints@@{\iflimtoken@                                              
 \def\ints@@@{\iflimits@\negintic@
   \mathop{\intic@\multintlimits@}\limits                             
  \else\multint@\nolimits\fi                                          
  \eat@}
 \else                                                                
 \def\ints@@@{\iflimits@\negintic@
  \mathop{\intic@\multintlimits@}\limits\else
  \multint@\nolimits\fi}\fi\ints@@@}%
\def\intkern@{\mathchoice{\!\!\!}{\!\!}{\!\!}{\!\!}}%
\def\plaincdots@{\mathinner{\cdotp\cdotp\cdotp}}%
\def\intdots@{\mathchoice{\plaincdots@}%
 {{\cdotp}\mkern1.5mu{\cdotp}\mkern1.5mu{\cdotp}}%
 {{\cdotp}\mkern1mu{\cdotp}\mkern1mu{\cdotp}}%
 {{\cdotp}\mkern1mu{\cdotp}\mkern1mu{\cdotp}}}%
\def\RIfM@{\relax\protect\ifmmode}
\def\text{\RIfM@\expandafter\text@\else\expandafter\mbox\fi}
\let\nfss@text\text
\def\text@#1{\mathchoice
   {\textdef@\displaystyle\f@size{#1}}%
   {\textdef@\textstyle\tf@size{\firstchoice@false #1}}%
   {\textdef@\textstyle\sf@size{\firstchoice@false #1}}%
   {\textdef@\textstyle \ssf@size{\firstchoice@false #1}}%
   \glb@settings}

\def\textdef@#1#2#3{\hbox{{%
                    \everymath{#1}%
                    \let\f@size#2\selectfont
                    #3}}}
\newif\iffirstchoice@
\firstchoice@true
\def\Let@{\relax\iffalse{\fi\let\\=\cr\iffalse}\fi}%
\def\vspace@{\def\vspace##1{\crcr\noalign{\vskip##1\relax}}}%
\def\multilimits@{\bgroup\vspace@\Let@
 \baselineskip\fontdimen10 \scriptfont\tw@
 \advance\baselineskip\fontdimen12 \scriptfont\tw@
 \lineskip\thr@@\fontdimen8 \scriptfont\thr@@
 \lineskiplimit\lineskip
 \vbox\bgroup\ialign\bgroup\hfil$\m@th\scriptstyle{##}$\hfil\crcr}%
\def\Sb{_\multilimits@}%
\def\endSb{\crcr\egroup\egroup\egroup}%
\def\Sp{^\multilimits@}%

\newdimen\ex@
\def\rightarrowfill@#1{$#1\m@th\mathord-\mkern-6mu\cleaders
 \hbox{$#1\mkern-2mu\mathord-\mkern-2mu$}\hfill
 \mkern-6mu\mathord\rightarrow$}%
\def\leftarrowfill@#1{$#1\m@th\mathord\leftarrow\mkern-6mu\cleaders
 \hbox{$#1\mkern-2mu\mathord-\mkern-2mu$}\hfill\mkern-6mu\mathord-$}%
\def\leftrightarrowfill@#1{$#1\m@th\mathord\leftarrow
\mkern-6mu\cleaders
 \hbox{$#1\mkern-2mu\mathord-\mkern-2mu$}\hfill
 \mkern-6mu\mathord\rightarrow$}%
\def\overrightarrow{\mathpalette\overrightarrow@}%
\def\overrightarrow@#1#2{\vbox{\ialign{##\crcr\rightarrowfill@#1\crcr
 \noalign{\kern-\ex@\nointerlineskip}$\m@th\hfil#1#2\hfil$\crcr}}}%

\def\overleftarrow{\mathpalette\overleftarrow@}%
\def\overleftarrow@#1#2{\vbox{\ialign{##\crcr\leftarrowfill@#1\crcr
 \noalign{\kern-\ex@\nointerlineskip}$\m@th\hfil#1#2\hfil$\crcr}}}%
\def\overleftrightarrow{\mathpalette\overleftrightarrow@}%
\def\overleftrightarrow@#1#2{\vbox{\ialign{##\crcr
   \leftrightarrowfill@#1\crcr
 \noalign{\kern-\ex@\nointerlineskip}$\m@th\hfil#1#2\hfil$\crcr}}}%
\def\underrightarrow{\mathpalette\underrightarrow@}%
\def\underrightarrow@#1#2{\vtop{\ialign{##\crcr$\m@th\hfil#1#2\hfil
  $\crcr\noalign{\nointerlineskip}\rightarrowfill@#1\crcr}}}%

\def\underleftarrow{\mathpalette\underleftarrow@}%
\def\underleftarrow@#1#2{\vtop{\ialign{##\crcr$\m@th\hfil#1#2\hfil
  $\crcr\noalign{\nointerlineskip}\leftarrowfill@#1\crcr}}}%
\def\underleftrightarrow{\mathpalette\underleftrightarrow@}%
\def\underleftrightarrow@#1#2{\vtop{\ialign{##\crcr$\m@th
  \hfil#1#2\hfil$\crcr
 \noalign{\nointerlineskip}\leftrightarrowfill@#1\crcr}}}%

\def\qopnamewl@#1{\mathop{\operator@font#1}\nlimits@}
\let\nlimits@\displaylimits
\def\setboxz@h{\setbox\z@\hbox}

\def\varlim@#1#2{\mathop{\vtop{\ialign{##\crcr
 \hfil$#1\m@th\operator@font lim$\hfil\crcr
 \noalign{\nointerlineskip}#2#1\crcr
 \noalign{\nointerlineskip\kern-\ex@}\crcr}}}}

 \def\rightarrowfill@#1{\m@th\setboxz@h{$#1-$}\ht\z@\z@
  $#1\copy\z@\mkern-6mu\cleaders
  \hbox{$#1\mkern-2mu\box\z@\mkern-2mu$}\hfill
  \mkern-6mu\mathord\rightarrow$}
\def\leftarrowfill@#1{\m@th\setboxz@h{$#1-$}\ht\z@\z@
  $#1\mathord\leftarrow\mkern-6mu\cleaders
  \hbox{$#1\mkern-2mu\copy\z@\mkern-2mu$}\hfill
  \mkern-6mu\box\z@$}

\def\projlim{\qopnamewl@{proj\,lim}}
\def\injlim{\qopnamewl@{inj\,lim}}
\def\varinjlim{\mathpalette\varlim@\rightarrowfill@}
\def\varprojlim{\mathpalette\varlim@\leftarrowfill@}
\def\varliminf{\mathpalette\varliminf@{}}
\def\varliminf@#1{\mathop{\underline{\vrule\@depth.2\ex@\@width\z@
   \hbox{$#1\m@th\operator@font lim$}}}}
\def\varlimsup{\mathpalette\varlimsup@{}}
\def\varlimsup@#1{\mathop{\overline
  {\hbox{$#1\m@th\operator@font lim$}}}}

\begingroup \catcode `|=0 \catcode `[= 1
\catcode`]=2 \catcode `\{=12 \catcode `\}=12
\catcode`\\=12 
|gdef|@alignverbatim#1\end{align}[#1|end[align]]
|gdef|@salignverbatim#1\end{align*}[#1|end[align*]]

|gdef|@alignatverbatim#1\end{alignat}[#1|end[alignat]]
|gdef|@salignatverbatim#1\end{alignat*}[#1|end[alignat*]]

|gdef|@xalignatverbatim#1\end{xalignat}[#1|end[xalignat]]
|gdef|@sxalignatverbatim#1\end{xalignat*}[#1|end[xalignat*]]

|gdef|@gatherverbatim#1\end{gather}[#1|end[gather]]
|gdef|@sgatherverbatim#1\end{gather*}[#1|end[gather*]]

|gdef|@gatherverbatim#1\end{gather}[#1|end[gather]]
|gdef|@sgatherverbatim#1\end{gather*}[#1|end[gather*]]

|gdef|@multilineverbatim#1\end{multiline}[#1|end[multiline]]
|gdef|@smultilineverbatim#1\end{multiline*}[#1|end[multiline*]]

|gdef|@arraxverbatim#1\end{arrax}[#1|end[arrax]]
|gdef|@sarraxverbatim#1\end{arrax*}[#1|end[arrax*]]

|gdef|@tabulaxverbatim#1\end{tabulax}[#1|end[tabulax]]
|gdef|@stabulaxverbatim#1\end{tabulax*}[#1|end[tabulax*]]

|endgroup

\def\align{\@verbatim \frenchspacing\@vobeyspaces \@alignverbatim
You are using the "align" environment in a style in which it is not defined.}

\@namedef{align*}{\@verbatim\@salignverbatim
You are using the "align*" environment in a style in which it is not defined.}
\expandafter\let\csname endalign*\endcsname =\endtrivlist

\def\alignat{\@verbatim \frenchspacing\@vobeyspaces \@alignatverbatim
You are using the "alignat" environment in a style in which it is not defined.}

\@namedef{alignat*}{\@verbatim\@salignatverbatim
You are using the "alignat*" environment in a style in which it is not defined.}
\expandafter\let\csname endalignat*\endcsname =\endtrivlist

\def\xalignat{\@verbatim \frenchspacing\@vobeyspaces \@xalignatverbatim
You are using the "xalignat" environment in a style in which it is not defined.}

\@namedef{xalignat*}{\@verbatim\@sxalignatverbatim
You are using the "xalignat*" environment in a style in which it is not defined.}
\expandafter\let\csname endxalignat*\endcsname =\endtrivlist

\def\gather{\@verbatim \frenchspacing\@vobeyspaces \@gatherverbatim
You are using the "gather" environment in a style in which it is not defined.}

\@namedef{gather*}{\@verbatim\@sgatherverbatim
You are using the "gather*" environment in a style in which it is not defined.}
\expandafter\let\csname endgather*\endcsname =\endtrivlist

\def\multiline{\@verbatim \frenchspacing\@vobeyspaces \@multilineverbatim
You are using the "multiline" environment in a style in which it is not defined.}

\@namedef{multiline*}{\@verbatim\@smultilineverbatim
You are using the "multiline*" environment in a style in which it is not defined.}
\expandafter\let\csname endmultiline*\endcsname =\endtrivlist

\def\arrax{\@verbatim \frenchspacing\@vobeyspaces \@arraxverbatim
You are using a type of "array" construct that is only allowed in AmS-LaTeX.}

\def\tabulax{\@verbatim \frenchspacing\@vobeyspaces \@tabulaxverbatim
You are using a type of "tabular" construct that is only allowed in AmS-LaTeX.}

\@namedef{arrax*}{\@verbatim\@sarraxverbatim
You are using a type of "array*" construct that is only allowed in AmS-LaTeX.}
\expandafter\let\csname endarrax*\endcsname =\endtrivlist

\@namedef{tabulax*}{\@verbatim\@stabulaxverbatim
You are using a type of "tabular*" construct that is only allowed in AmS-LaTeX.}
\expandafter\let\csname endtabulax*\endcsname =\endtrivlist

\def\@@eqncr{\let\@tempa\relax
    \ifcase\@eqcnt \def\@tempa{& & &}\or \def\@tempa{& &}%
      \else \def\@tempa{&}\fi
     \@tempa
     \if@eqnsw
        \iftag@
           \@taggnum
        \else
           \@eqnnum\stepcounter{equation}%
        \fi
     \fi
     \global\tag@false
     \global\@eqnswtrue
     \global\@eqcnt\z@\cr}

 \def\endequation{%
     \ifmmode\ifinner 
      \iftag@
        \addtocounter{equation}{-1} 
        $\hfil
           \displaywidth\linewidth\@taggnum\egroup \endtrivlist
        \global\tag@false
        \global\@ignoretrue   
      \else
        $\hfil
           \displaywidth\linewidth\@eqnnum\egroup \endtrivlist
        \global\tag@false
        \global\@ignoretrue 
      \fi
     \else   
      \iftag@
        \addtocounter{equation}{-1} 
        \eqno \hbox{\@taggnum}
        \global\tag@false%
        $$\global\@ignoretrue
      \else
        \eqno \hbox{\@eqnnum}
        $$\global\@ignoretrue
      \fi
     \fi\fi
 } 

 \newif\iftag@ \tag@false
 
 \def\tag{\@ifnextchar*{\@tagstar}{\@tag}}
 \def\@tag#1{%
     \global\tag@true
     \global\def\@taggnum{(#1)}}
 \def\@tagstar*#1{%
     \global\tag@true
     \global\def\@taggnum{#1}%
}

\begin{document}

\author{Bert Schroer \\
Freie Universit\"at Berlin\\
Institut f\"ur theoretische Physik\\
Arnimalle 14 14195 Berlin\\
e-mail schroer@physik.fu-berlin.de}
\title{Wigner Representation Theory of the Poincar\'e Group, Localization ,
Statistics and the S-Matrix }
\date{June 1996 }
\maketitle

\begin{abstract}
It has been known that the Wigner representation theory for positive energy
orbits permits a useful localization concept in terms of certain lattices of
real subspaces of the complex Hilbert -space. This ''modular localization''
is not only useful in order to construct interaction-free nets of local
algebras without using non-unique ''free field coordinates'', but also
permits the study of properties of localization and braid-group statistics
in low-dimensional QFT. It also sheds some light on the string-like
localization properties of the 1939 Wigner's ''continuous spin''
representations.We formulate a constructive nonperturbative program to
introduce interactions into such an approach based on the Tomita-Takesaki
modular theory. The new aspect is the deep relation of the latter with the
scattering operator.
\end{abstract}

\section{ Introduction.}

The main aim of this paper is the exploration and extension of Wigner's
representation-theoretic approach to relativistic quantum theory \cite
{Wigner} for the construction of particles and free fields in the context of
d=2+1 abelian braid-group statistics (the particles are often referred to as
''anyons or ''abelian plektons ''). In this way one may hope to obtain a
more direct understanding of the origin and the physical consequences of
plektonic statistics than that by the somewhat vague (and often indirect and
complicated) method via imposing Chern-Simons perturbation on standard
fermionic matter, using the formalism of functional integrals (in a region
where the necessary and sufficient conditions for Feynman-Kac
representability of quantum physics are strictly speaking violated).

The main issue in the adaptation of Wigner's theory to this question is the
problem of ''localization''. The sharp and covariant concept used in these
notes is not that of Newton and Wigner \cite{Newton}, but the more recent
idea of localization via suitably defined real subspaces \cite{Leyland} of
the complex Wigner representation space. This concept relates to the rather
universal and deep mathematical Tomita-Takesaki modular theory for von
Neumann algebras \cite{Tomita} which has been known to connect to such
diverse looking physical issues as the existence of antiparticles (TCP), the
stability of temperature states, the Unruh-Hawking effect \cite{Sewell} and
many other structures of ''Local Quantum Physics ''\cite{Haag}. The
underlying ''modular'' wedge-localization unlike the standard tools (as e.g.
the Gell-Man-Low perturbation theory in terms of time-ordered products or
the functional integral approach) has no counterpart in classical field
theory or in nonrelativistic quantum theory. {\it Among all concepts in QFT
it is the most intrinsic one, }and it achieves something which in e.g. in
coordinate-free differential geometry was accomplished already a long time
ago, namely the separation of intrinsic properties from mainly accidental
''coordinatizations''.

In section 2 entitled ''ancient history '' we review those aspects of
Wigner's theory which are relevant for our purpose. This theory was the
first successful attempt to formulate a framework of relativistic quantum
theory {\it not based on the quantization} parallelism to classical
theories. Mainly through algebraic QFT, its spirit of only using
intrinsically defined concepts has been kept alive in present day field
theory. To most physicists of the younger generation who are familiar with
the perturbative aspects of the Lagrangian quantization approach, the Wigner
theory remained largely unknown because modern texts often equate QFT with
the Lagrangian approach and path integrals.

Besides the (perhaps too brief on this issue) reference \cite{Haag} , the
reader can find some material in S.Weinberg's recent book \cite{Weinberg}.
But in the latter the Wigner approach is unfortunately (against Wigner's
intention to use only intrinsic concepts of quantum physics) mainly used in
order to {\it support} the Lagrangian formulation of QFT and the
path-integral approach (although the possibility of other non-Lagragian
approaches to interactions is at least not ruled out, as emphasized by the
author). Here our aim is quite different, namely to understand areas which
are not covered (and probably never will be) by the perturbative Lagrangian
framework. It is worthwhile to mention that all the pertinent results on
chiral conformal QFT as well as a large part of results on massive d=1+1
theories have been obtained by nonperturbative non-Lagrangian methods such
as representation theory, S-matrix bootstrap, formfactor program etc. A
Lagrangian name, where it appears, usually only served for ''baptizing'' the
model\footnote{%
In the standard approach the use of different field coordinates has
sometimes caused an inflatory use of names viz. the
Wess-Zumino-Witten-Novikov model which is just the generalized Thirring
model at the critical point in the multicomponent coupling constant space
rewritten in different field coordinates which are more adapted to the
differential geometric mathematical exploitation than the Thirring
parametrization (which is very useful in condensed matter physics).} in the
traditional way but plays no role in its construction.

In fact the fields which appeared in the 1974 work on conformal field theory 
\cite{swieca} were so remote from Lagrangian-and Euclidean- (and even from
Wightman-) fields, that the problem of a systematic model construction was
not pursued as a result of Zeitgeist prejudices, which pointed into the
direction of euclidean theory. Nowadays it is very natural to consider
charged fields which have nontrivial source and range projectors onto
superselection sectors, but at that time this appeared as going against the
holy gral of QFT.

In fact one would be surprised if plektonic d=2+1 fields or those
corresponding to Wigner's d=3+1 ''continuous spin'' representations are not
of this new{\it \ non-Lagrangian} kind. An educated guess is that all fields
in $d\geq 2+1$ which create only states with weaker than compact
localization properties are non-Lagrangian. This is in agreement with an old
result of Yngvason about the obstructions posed by the d=3+1, m=0 Wigner
''continuous spin'' representation within a Wightman framework \cite{Mack}.

In section 3 entitled ''recent history '', we briefly present those aspects
of the Tomita Takesaki modular theory which are relevant in the present
context. In particular we explain, how by introducing {\it real} subspaces
of the Wigner representation space via a Tomita involution, one may
implement a localization concept which is more useful for our purpose than
the Newton-Wigner localization. These ideas, although known to some experts,
unfortunately had never been published in an accessible way\cite{Leyland}.
Section 3 also contains a brief sketch of a direct construction of local
nets from the so called ''wedge localization''\cite{Bisognano-Wichmann}
which, in the case of free bosonic theories, will be treated in more details
and presented within a more general mathematical setting in forthcoming work
of Brunetti, Guido and Longo\cite{brun}.

The fourth section explains why the adaptation of Wigner's theory for d=2+1
anyonic spin is not compactly localizable, but still falls into the weaker
spacelike cone-(or semi-infinite string-) localizable category and presents
the corresponding ''anyonic'' statistics in terms of a ''twist ''which is
necessary in order to balance the dual quantum localization of the
wedge-localized real subspace (i.e. its symplectic complement) with the
geometric (causal) dual in the sense of Lorentz-transformations relations of
fields.

In section 5, we comment on a an interesting topological obstruction against
Haag duality for non-simply connected regions which occurs in certain zero
mass theories including the free Maxwell theory. These topological
obstructions are absent in massive theories, but they are typical of local
gauge theories (i.e. those Lagrangian theories for which long range
interaction can presently only be described by first introducing an
indefinite metric vector potential and a formal return to quantum physics
via a perturbative BRST condition). Algebraic QFT was not able (for very
good reasons in my opinion) to incorporate classical gauge ideas which have
their natural formulation in fibre bundle theory. But through such duality
obstructions as discussed in section 5, the algebraic approach at least
perceives that there is a deep problem on the level of local quantum physics
for a certain type of theories involving zero mass. In my view an adequate
treatment of this problem can only be given in a framework of interaction
which uses concepts which are characteristic of relativistic QFT as e.g. the
modular properties used in this work.

In section 6 we show that Wigner's zero mass ''continuous spin''
representation falls into this weaker space-like cone localization category.
In fact the natural covariant description \cite{Mack} is in terms of
semi-infinite light-like strings, very similar to the covariantization
attempt of the d=2+1 ''Wigner anyons'' mentioned in section 4.

The last section contains some speculative attempts of incorporating an
intrinsic notion of interaction via an ansatz for an ''interacting'' Tomita
involution $J$ within the Fock-space setting defined by scattering theory.
This Ansatz generalizes the $J_{0}$ obtained from the Wigner theory and does
not involve the interaction picture and time-ordering, but uses only
nonperturbative concepts of general quantum field theory. In this last
section I also speculate on several presently insufficiently understood
problems which have connections with modular ideas.

\section{Ancient History.}

In 1939 Wigner \cite{Wigner} classified the irreducible ray-representations
of the Poincar\'{e}-group (or what amounts to the same, the irreducible
vector-representations of its covering). His main motivation was to
understand in intrinsic physical terms the ever increasing ``zoo'' of linear
relativistic (higher spin) field equations of those days, which were
proposed in the aftermath of the Dirac equation. For this purpose he had to
extend the Frobenius method of induced representations from finite groups to
the non-compact Poincar\'{e}-group, a mathematical novelty which gave rise
to mathematical developments in group representations, \cite{Bargmann}. He
first determined all transitive momentum-space orbits under the
Lorentz-group and then classified the (isomorphic for different momenta on
the same orbit) fixpoint-group of a conveniently chosen reference vector $%
p_{R}$ on the orbit. The ``induction'' was done with the help of this
``little group'' and a suitably defined family of ``boosts'' served to
identify the fixpoint-groups at different orbit points.

For the positive energy orbits $p^2=m^2,p_0>0$ and $p^2=0,p_0>0$ (the only
orbits of relevance for our purpose) in d=3+1, the (coverings of the) little
groups are $SU(2)\,$ resp. $\widetilde{E}^{\left( 2\right) }(2)$ (the
two-fold covering of the two-dimensional Euclidean group).

The massive $\left[ m,s\right] $-representations are most conveniently
described in terms of 2s+1 component wave-function spaces: 
\begin{equation}  \label{Eq.1}
H=\left\{ \underline{\psi }(p)\right| \sum_{s_3}\int \left| \psi
_{s_3}(p)\right| ^2\frac{d^3p}{2\omega }<\infty \}
\end{equation}
on which the Lorentz transformation acts as: 
\begin{equation}  \label{Eq.2}
(U(\Lambda )\underline{\psi })(p)=D^{(s)}(R_W(\Lambda ,p)) \underline{\psi }%
(\Lambda ^{-1}p)
\end{equation}
with R$_W$ being the $\Lambda -$ and p-dependent (nonlocal) Wigner rotation.

In the $m^{2}=0$ case one has a greater wealth for the representation theory
of the little group. In case the ''translations ''of $\widetilde{E}(2)$ are
mapped to zero, one obtains the family of nonfaithful one-component
semi-integers-helicity representations:

\begin{equation}  \label{eq.3}
(U(\Lambda )\psi )(p)=e^{is\Phi _W(\Lambda ,p)}\psi (\Lambda ^{-1}p)
\end{equation}

The conversion of the one-component Wigner wave functions into e.g. the
standard local helicity description in terms of field strength $F_{\mu \nu }$
is well- known .

Explicit formulas for the Wigner phase $\Phi _{W}$ as well as the previous
Wigner rotation $R_{W}$ are to be found already in the original paper as
well as in S. Weinberg's recent book.\cite{Weinberg} Also the extensions to
the full group including space and time reflections may be found in the
literature. In the following we will need the formula for the $TCP=\theta $
transformations acting on the (doubled, if particles are not self-conjugate) 
$\left[ m,s\right] $-representation as:

\begin{equation}
\theta \left( 
\begin{array}{c}
\psi _{+} \\ 
\psi _{-}
\end{array}
\right) =D^{(s)}(i\sigma _{2})\left( 
\begin{array}{c}
\psi _{-}^{*} \\ 
\psi _{+}^{*}
\end{array}
\right) ,\,\,\,\,\,\pm \mbox{=(anti)particle\thinspace \thinspace doubling}
\label{Equ.4}
\end{equation}

Before we relate this TCP-transformation of the Wigner theory to a new
localization concept, some more historical remarks are in order.

Wigner's work, although little noticed at the time (at least by the
community of producers of new relativistic field equations), showed in one
stroke that the problem of inventing more general looking field equations
was of a somewhat academic nature; what really mattered for the particle
content was their irreducible p-space representation structure, and not
their covariant appearance in x-space.

Wigner was apparently aware that Poincar\'{e}-invariance was not the only
physical requirement for relativistic particles, but there were also the
important issues of causality and localization. In 1949 he wrote a paper
together with R.Newton \cite{Newton} in which they proposed, what became
later known as the Newton-Wigner localization .This localization was not
covariant and effectively violated Einstein causality at distances shorter
than a Compton-wavelength, but it seemed to be the best one could do if one
adapts the wave-packet localization of the Schroedinger theory to the
relativistic domain.

As a result of these unsatisfactory aspects of this localization, Wigner
became increasingly suspicious about the internal consistency of QFT
(private remark obtained from R.Haag). However a short time later Wightman
and collaborators showed that there was no contradiction between the
Heisenberg-Pauli canonical quantization approach and the Wigner theory\cite
{Wightman}. In fact the latter can be used in order to obtain a more
intrinsic access to the former \cite{Weinberg}.

With one $\left[ m,s\right] $-representation one connects a whole family of
free fields which all share the same canonical momentum space creation and
annihilation operators affiliated (transforming) with the $\left[ m,s\right] 
$ Wigner representation:

\begin{equation}  \label{Equ.5}
\Psi (x)=\frac 1{(2\pi )^{\frac 32}}\int \left(
e^{-ipx}\sum_{s_3=-s}^su(p,s_3)a(p,s_3)+e^{ipx}%
\sum_{s_3=-s}^sv(p,s_3)b^{*}(p,s_3)\right) \frac{d^3p}{2\omega }
\end{equation}

Here u and v are explicitly known column-vectors in a space of $\geq 2s+1$
dimensions. They represent intertwiners between the Wigner representation
and the covariant description of its content:

\begin{equation}
\sum_{s_{3}^{\prime }}u(p,s_{3}^{\prime })D_{s_{3},s_{3^{\prime
}}}^{(s)}(R_{W}(\Lambda ,p))=D_{covar}^{\left[ n,m\right] }(\Lambda
)u(\Lambda ^{-1}p,s_{3})  \label{Equ.6}
\end{equation}

The indexing of the entries of u is given by a pair $(n,m)$ of n un-dotted
and m dotted symmetrised spinorial indices ($\alpha _{1}\alpha _{2}$ ....$%
\alpha _{n},\beta _{1}\beta _{2}.....\beta _{m}$). The only restrictions are
that $\frac{n+m}{2}$ be semi-integer if s is semi-integer as well as the
validity of the inequality $\left| \frac{n}{2}-\frac{m}{2}\right| \leq s\leq 
\frac{n}{2}+\frac{m}{2}.$ Hence the matrix $D_{covar}$ describes a
finite-dimensional tensorial (and therefore non-unitary) representations of
the Lorentz-group. A systematic determination of this infinitely large
family of intertwiners for fixed $\left[ m,s\right] $ is not contained in
the original work, but was carried out later by Joos\cite{Joos} and
Weinberg. In Weinberg's recent book \cite{Weinberg} the reader finds an
exhaustive treatment of this family of local fields which all share the same
momentum space creation and annihilation operators. There one also finds a
careful discussion of some peculiarities of the $\left[ 0,h\right] $
photon-neutrino class. In that case the covariantization of these
nonfaithfull Wigner representation is much more restrictive than for massive
theories. Whereas for the latter case one has the above inequality , the
helicity for the former obeys the equality $h=\left| \frac{n}{2}-\frac{m}{2}%
\right| $. For the much more elusive infinite component ''continuous spin''
faithful zero mass representation, the covariantization was carried out in
the 70's where also the lack of the standard localization property was
noticed \cite{Mack} but the modular concepts for a sharp localization
investigation were not yet available.

In all cases whether massive or zero mass, the local covariant fields live
in the same Fock-space i.e they share the same momentum space creation and
annihilation operators and in addition are local relative to each other in
the sense of space-like (anti)commutation relations. Using a very
appropriate concept of Borchers \cite{Borchers}, one obtains a more concise
description of this notion of relative locality. Namely it turns out that
these fields are members of an equivalence class of relatively local fields.
More specifically, they form a linear subset of the free field $\left[
m,s\right] $ '' Borchers class '', an object which has been explicitly
computed in the 60's by H.Epstein and the present author\cite{Streater}%
\footnote{%
It consists precisely of the Wick-ordered composites including derivatives.}%
. Borchers showed in complete generality that fields, which are local with
respect to a given local field, with the latter acting cyclically on a
Hilbertspace, are automatically local (with respect to themselves) and he
proved that this entails the following consequences:

\begin{itemize}
\item  (i) The cyclically acting members generate the same local von
Neumann-algebras, i.e. if $A(x)$ and $B(x)$ are two such fields and ${\cal A}%
(A,{\cal O})$ denotes the local von Neumann-algebra generated by the field
A(x) smeared with test-functions having support in ${\cal O}$ (a natural
family of regions ${\cal O}$ are the so called double cones on which
Poincare-transformations act stably) one has:
\end{itemize}

\begin{equation}
{\cal A}(A,{\cal O})={\cal A}(B;{\cal O})  \label{Equ.7}
\end{equation}

\begin{itemize}
\item  (ii) The different members of the Borchers class do not only lead to
the same local observables, but also entail the same S-matrix i.e. the
S-matrix is a class invariant.
\end{itemize}

This suggests a viewpoint of QFT ( ''algebraic QFT '') which is quite
different from the standard one in most of the textbooks, although it is
based on the same physical principles. By analogy with differential
geometry, the pointlike covariant fields are like coordinates and the
algebraic net, i.e.the assignment: ${\cal O}\rightarrow A({\cal O})$
contains all the intrinsic physical information \cite{Haag}. The terminology
''field coordinates'', which is used freely in the present work, is
precisely meant in this sense. The Borchers theory also gave the prominent
role as a net invariant to the S-matrix. In the last section we will use the
S-matrix as an invariant of the wedge-based modular theory.

Progress obtained from this net point of view has been slow, but steady and
very solid indeed. The accumulated body of results is quite impressive by
now. Its mathematical pillars are the Tomita-Takesaki modular theory\cite
{Tomita} and the V.Jones subfactor theory\cite{Jones}, both dealing with
structural properties of von Neumann-algebras. It is not an accident, that
both mathematical theories had their physical (mathematically less general)
predecessors: the Haag-Hugenholtz-Winnink \cite{Haag} description of
KMS-states in the first case, and the Doplicher-Haag-Roberts superselection
theory \cite{Haag} in the second (a fact which was not known at the time of
the mathematical discoveries).

Here we want to show that the seeds for this intrinsic mode of physical
thinking are already contained in the Wigner theory. More concretely, the
Wigner theory preempts some special aspects of both mathematical theories:
localization properties are related to modular properties (explained in
detail in the sequel) and duality obstructions related to properties of
inclusions (briefly mentioned in section 5). It is interesting to note that
this progress occurs precisely at the localization structure which Wigner
considered questionable.

Needless to add the remark that the $\left[ m,s\right] $ fields are in
general not Lagrangian fields i.e. the above local free fields are in
generally not solutions of an Euler-Lagrange equation. To give an example,
for $s=\frac{3}{2}$ the Rarita-Schwinger field is ''Lagrangian '', but the
e.g. minimal 4-component field in the same Borchers-class is not
''Eulerian'', i.e. its Lorentz transformation properties cannot be
incorporated into the structure of an Euler-Lagrange field equations,
neither are those fields in the range of the canonical formalism which is an
important property of the Lagrangian field theory and the Cauchy initial
value problem. Euler-Lagrange structures are not relevant for {\it quantum}
field theory and the fact that each free field Borchers class contains one
Euler-Lagrange representative does not seem to bring about any additional
insight. With the modular structures we will even move further away from
classical properties and quantization prescriptions.

In Weinbergs book one finds a formal argument which indicate that invariant
(Wick-ordered) polynomial coupling terms lead to perturbations which are
independent of the $\left[ m,s\right] $ field-coordinates which one uses for
the specification of the interaction density. With other words, a given
polynomial interaction may be rewritten in terms of any kind of
field-coordinates one likes (and it stays polynomial in terms of the new
fields). This observation suggests that even in a perturbative approach one
should try to avoid these ''field-coordinates'' altogether and aim for a
description which restores Wigner's representation uniqueness on the level
of the associated operator algebras. In this way one could avoid the
confusing multitude of field coordinates and hopefully eventually also
arriving at a more intrinsic understanding of interactions. The next section
takes a step towards this goal.

\section{Recent History}

For the sake of simplicity let us assume that we are using the Wigner
formalism in order to describe a self-conjugate particle situation.Then,
apart from a possible sign factor, the previous action of $\theta =$ $TCP$
on the momentum space wave-function simplifies as follows: 
\begin{equation}
(\theta \psi )(p)=D^{(s)}(i\sigma _{2})\psi ^{*}(p)  \label{Equ.8}
\end{equation}
where $D^{(s)}(i\sigma _{2})$ represents the conjugation matrix in the $%
\left[ m,s\right] $ Wigner space. Introduce now another conjugation j which
differs from $\theta $ by a rotation with $\pi $ around the x-axis :

\begin{equation}  \label{Equ.9}
j=R(e_x,\pi )\cdot \theta
\end{equation}

An elementary calculation shows that this $j$ commutes with the L-boost $%
\Lambda _{W_{st}}(\chi )$ assciated to the standard wedge which we have
chosen in the 0-1 -plane :

\begin{equation}
U(\Lambda _{W_{st}}(\chi ))j=jU(\Lambda _{W_{st}}(\chi ))  \label{Equ.10}
\end{equation}

Whereas the boosts define a unitary subgroup, the continuation to imaginary $%
\chi $ yields an unbounded closable operator:

\begin{equation}
\delta :=U(\Lambda (\chi =2\pi i))  \label{Equ.11}
\end{equation}

In particular the unbounded operator: 
\begin{equation}
s_{Wst}=j\delta ^{\frac{1}{2}}  \label{Equ.12}
\end{equation}
turns out to be a closed densely defined antilinear involution. This is as a
result of the commutation relation :

\begin{equation}  \label{Equ.13}
j\delta ^{\frac 12}=\delta ^{-\frac 12}j
\end{equation}
which follows from the one before.

In fact the above definition of $s_{W_{st}}$ agrees with its polar
decomposition. We now use this involution $s$ in order to define a closed
real subspace : 
\[
H_{W_{st}}^{R}=\{h\in H\;\mid {\sl s}_{W_{st}}h=h\} 
\]

The properties of the positive operator $\delta $ entail the density of $%
H_{W_{st}}^{R}+iH_{W_{st}}^{R}$ in the Wigner representation space $.$ This
decomposition also allows to introduce a real inner product on $H$ : 
\begin{equation}
(\psi ^{\prime }+i\chi ^{\prime },\psi +i\chi )^{R}=(\psi ^{\prime },\psi
)+(\chi ^{\prime },\chi )  \label{equ.14a}
\end{equation}
where the primed wave-functions belong to the (-1) eigenspace $%
H_{W_{st}}^{R\prime }$ of the adjoint $s{\sl _{Wst}^{*}}$. Both summands on
the right hand side are real since:

\begin{equation}
(\psi ^{\prime },\psi )=-(\psi ^{\prime },s_{W_{st}}\psi )=-\overline{%
(s_{W_{st}}^{*}\psi ^{\prime },\psi )}=-\left( \psi ,s_{W_{st}}{\sl ^{*}}%
\psi ^{^{\prime }}\right) =(\psi ,\psi ^{\prime })  \label{Equ.14b}
\end{equation}

The real structure is the same as the one obtained by using the real part of
the complex inner product and then restricting to the subspace $%
H_{W_{st}}^{R}$. Conversely one can obtain $H_{W_{st}}$ by introducing a
complex structure on $H_{W_{st}}^{R}.$ By applying Poincar\'{e}
transformations to those spaces, one obtains a whole family of real
subspaces which are eigenspaces of densely defined involutions $s_{W}$
corresponding to the family of wedges obtained from the t-x wedge by
Poincare-transformations. One then finds the following surprising theorem :

{\it Theorem(Brunetti,Guido and Longo\cite{brun}): The family of real wedge
subspaces form a covariant net of wedge-localized subspaces.}

This means in particular that one has isotony i.e.$H_{W^{^{\prime
}}}^{R}\subset H_{W}^{R}$ for $W^{^{\prime }}\subset W$ and it is
interesting to note that this inclusion property is equivalent to the
positivity of the energy {\it \cite{brun}}

It is well known that in case of integer Wigner spin there exists the so
called Weyl functor \cite{Leyland} which converts these localized real
subspaces into local von Neumann algebras. The generators of these algebras
in physicists notation are: 
\begin{equation}
W(h)=e^{i(\Psi (h)+\Psi (h)^{*})}\,\,,\,\,\,\,\,h\in H_{W}^{R}
\label{equ.15}
\end{equation}

In other words the algebras for the wedge regions can be directly defined in
terms of the Wigner theory without reference to ''local field coordinates''.
Algebras of e.g.double cones may be formed through intersections. Their
associated real subspaces lead to complexifications which are dense in the
Wigner space (apart from the ''continuous spin'' representation). This can
be seen by using localization properties of the u,v intertwiners of the
previous section. A more elegant way, which presently is in the process of
being worked out, would be to isolate a property of the wedge subspaces
which guarantees this density without using any covariantizing intertwiners.

Physicists familiar with another ''miracle '' from the quantum physics in
curved space-time (for a recent review see\cite{Brout}) namely the
Unruh-Hawking effect for the Rindler wedge i.e. the quantum physics of a
uniformly accelerated observer, should take notice that this effect and the
above theorem are two sides of the same coin. Behind both miracles lies a
very basic and universal theory \cite{Sewell} which, as already mentioned in
the introduction, mathematicians refer to as the Tomita-Takesaki modular
theory. Physically one expects among other things that this theory explains
basic features of QFT as e.g. crossing symmetry of the S-matrix and
formfactors to be derivable from the KMS-temperature properties of the
Hawking-Unruh situation. We will here only limit ourselves to some salient
features. This theory deals with von Neumann-algebras in ''standard position
'' e. g. weakly closed operator algebras in a Hilbertspace possessing a
cyclic and separating vector $\Omega $. In local quantum physics the vacuum
is a vector which has this property with respect to all local subalgebras
with a nontrivial causal complement of their localization region \cite{Haag}%
. In order to construct the basic objects of this theory, one starts from
the *-structure of the von Neumann-algebra {\cal A} and defines an unbounded
but closable involutive operator S : 
\begin{equation}
SA\Omega =A^{+}\Omega \;,\quad A\in {\cal A}  \label{Equ.17}
\end{equation}

Its polar decomposition: 
\begin{equation}
S=J\Delta ^{\frac{1}{2}}  \label{Equ.18}
\end{equation}
defines the Tomita conjugation $J$ and the modular operator $\Delta $ . The
latter gives rise to the modular automorphism $\sigma _{t}.$ The nontrivial
part of the T.-T. theorem is about the behaviour of these operators with
respect to the von Neumann-algebra {\cal A}: 
\begin{equation}
J({\cal A}):=J{\cal A}J={\cal A}^{\prime },\,\,\,\,\,\,\,{\cal A}^{\prime
}:commutant\,\,of\,\,{\cal A}.  \label{Equ.19}
\end{equation}

\begin{equation}
\sigma _{t}({\cal A})=\Delta ^{it}{\cal A}\Delta ^{-it}={\cal A},\quad
\sigma _{t}:\,modular\ automorphism\ of\,\,{\cal A}  \label{Equ.20}
\end{equation}
For a physicist, the K in $\Delta ^{it}=e^{itK}$ is like a generalized
Hamiltonian and $J$ is like a generalized TCP-operator of the pair ${\cal A}%
,\Omega .$ The only miracle as far as the application of this theory to the
local algebras of QFT is concerned, is that these modular quantities for the
pair {\cal A}(wedge)$\,,\Omega \,$(vacuum) become geometric:

\begin{equation}
\Delta ^{it}=U(\Lambda _{W_{st}}(\chi =2\pi t)  \label{Equ.21}
\end{equation}

\begin{equation}
J=\left\{ 
\begin{array}{c}
R(e_{x},\pi )\Theta ,\mbox{ s= integer} \\ 
{\cal K}R(e_{x},\pi )\Theta ,\mbox{ for~ s= semi-integer}
\end{array}
\right.  \label{Equ.23}
\end{equation}

Here {\cal K} is the well-known Klein-twist (not to be confused with the
closely related Jordan-Wigner transformation) for fermions: ${\cal K}=\frac{%
1+iV}{1+i}$ with $V=\exp i\pi N_{fermi}.$ In the integer spin case the
Wigner theory preempts this modular structure through the existence of the
previously introduced family of real subspaces $H_{W}^{R}$ which are
converted into algebras of a bosonic net via the Weyl functor. In the
semi-integer spin case the CAR-functor plays an analogous role. In the
Wigner theory the Fermi-statistics manifests itself through:

\begin{equation}  \label{Equ24}
jH_{wedge}^R\neq H_{oppositewedge}^R%
\mbox{ \thinspace
\thinspace \thinspace \thinspace for s=semi-integer}
\end{equation}

This mismatch is repaired on the level of the algebras by the above
Klein-twist {\cal K}: 
\begin{equation}  \label{Equ.25}
{\cal KF}(jH_{wedge}^R)={\cal F}(H_{opp.wedge}^R)
\end{equation}

Here {\cal F\ }is the CAR functor. {\cal K} restricted to the Wigner-space
of fermions is just a numerical factor $i$, which is precisely the
obstruction factor between $j$ and the $\pi -$rotation. So the Klein factor
just permits to express the Tomita $J$ in terms of geometrical objects. The
rest consists in applying the CCR resp. CAR functor which maps the net of
Hilbert spaces into the net of von Neumann algebras.

Note that all recent contributions of modular theory to the understanding
and construction of Borchers classes (including the present one) could have
been given two decades ago, ever after the prominent role of wedge algebras
was discovered by Bisognano and Wichmann. \cite{Bisognano-Wichmann}. But as
it often happens, conceptual gains need a longer time for mental digestion
than gains in formalism.

\section{Fractional Wigner-Spin and Statistics of Anyons.}

In d=2+1, the little group of a point on the forward mass shell is the
abelian U(1) and therefore the Wigner theory allows (at least a priori) for 
{\it any} value of spin, i.e.one expects ''anyons '' (the more restrictive
non-abelian plektons will only be mentioned at the end of this section ).
Using the methods of the previous section, and checking the prerequisites
for the existence of a TCP operation on the direct sum of
particle-antiparticle Wigner spaces, one again establishes the properties of
a family of real subspaces which can be associated with localization and
statistics properties of field theoretic two point functions. However the
difference between the modular complement $jH_{W}^{R}$ $=H_{W}^{R\prime }$%
and the geometric complement $H_{W^{^{\prime }}}^{R}=Rot(\pi )H_{W}^{R}$ is
bigger than in the previous fermionic theory i.e.the Klein transformation
which accounts for this difference is more complicated.

In order to keep the Klein-twist simple, let us imagine that we are dealing
with a $Z_{N}-$spin i.e. we assume that $s=\frac{1}{N}$ .Then the Klein
factor which corrects the mismatch between localization via commutativity
and the geometric localization turns out to be a suitable ''square root''%
\cite{Schroer} of the action of the $2\pi -$rotation in space of scattering
states 
\begin{equation}
K=\sum_{n}e^{-i\pi sn^{2}}P_{n}  \label{Equ.26}
\end{equation}
This is in agreement with the nonlinear composition of anyonic spin. These
kinematical facts suggest that the scattering space cannot have the tensor
product structure as for Bosons and Fermions. A physically relevant question
is: what is the a priory best possible localization of the anyonic algebras?
Certainly the field theoretic localization cannot be better than the modular
localization in the Wigner representation space. It turns out that a compact
localization as in the previous section is not possible, i.e. in $H_{W}^{R}$
there are no compactly localized wave functions. If such a wave function
would exist, one could perform a $2\pi -$rotation such that the support
remains inside one wedge for all angles, however the nontrivial phase
created by such a rotation contradicts its affiliation to the real subspace $%
H_{W}^{R}$.

>From the general structure of algebraic QFT we expect that the spectral gap
leads to a (arbitrarily thin) space-like cone localization. In d=2+1 only
genuine braid group statistics is able to exhaust this possibility, whereas
permutation group statistics resulting from semi-integer spin leads back to
the compact localization. Since the core line of a semi-infinite spacelike
cone is characterized by an initial point x and a spacelike unit direction
e, we expect a string-like localized wave function depending on x and e.

Starting from the Wigner wave function which transforms according to ($\psi$
is one-component) :

\begin{equation}
(U(g)\psi )(p)=e^{is\Phi _{W}(g,p)}\psi (\Lambda ^{-1}(g)p),\qquad g\in 
\widetilde{SO(2,1)}  \label{Equ.27}
\end{equation}
with $\Phi _{W}$ being the (nonlocal) Wigner phase. One looks for a
factorization into covariant factors in analogy with the u-v intertwiners of
the previous section. This is achieved by \cite{Gaberdiel} defining: 
\begin{equation}
\psi _{covar}(p,g)=F(L^{-1}(p)g)\psi (p)  \label{Equ.28}
\end{equation}
where any function F on $\widetilde{SO(2,1)}$ is acceptable as long as it
fulfills the equivariance law: 
\begin{equation}
F(\omega \cdot g)=e^{is\omega }F(g)\qquad \omega \in R\subset \widetilde{%
SO(2,1)}\quad g\in \widetilde{SO(2,1)}  \label{Equ.29}
\end{equation}
As a consequence we find the covariance law: 
\begin{equation}
(U(g^{\prime })\psi _{cov})(p,g)=\psi _{cov}(\Lambda ^{-1}(g^{\prime
})p,gg^{\prime })  \label{Equ.30}
\end{equation}
With this covariant wave function we now affiliate a Dirac state vector
which, as usual, is created from the vacuum: 
\begin{equation}
\left| p,g\right\rangle =a^{*}(p,g)\Omega  \label{Equ.31}
\end{equation}
It obeys the contragradient transformation law: 
\begin{equation}
U(g)a^{*}(p,g^{\prime })U(g)^{-1}=a^{*}(\Lambda (g),g^{\prime }g^{-1})
\label{Equ.32}
\end{equation}
Transforming to x-space fields (the subscript cov will be omitted in the
sequel): 
\begin{equation}
\psi (x,g)=\int \frac{d^{2}p}{2\omega }(e^{ipx}a(p,g)+e^{-ipx}a^{*}(p,g))
\label{Equ.33}
\end{equation}
one obtains the desired transformation law: 
\begin{equation}
U(g^{\prime })\psi (x,g)U(g^{\prime })^{-1}=\psi (\Lambda (g^{\prime
})x,g^{\prime }g)  \label{Equ.34}
\end{equation}
The two-point function is a quadratic expression in the function F: 
\begin{equation}
(\Omega ,\psi (x_{1,}g_{1})\psi (x_{2},g_{2})\Omega )=\int \frac{d^{2}p}{%
2\omega }e^{ip(x_{1}-x_{2})}\overline{F(L^{-1}(p)g_{1})}\cdot
F(L^{-1}(p)g_{2})  \label{Equ.35}
\end{equation}
Choosing $x_{2}$ and $g_{2}\mbox{"opposite"}$ to $x_{1}$and $g_{2}$ i.e.
such that: 
\begin{equation}
x_{2}=Rot(\pi )x_{1\qquad }g_{2}=Rot(\pi )g_{1}  \label{Equ.36}
\end{equation}
the covariant transformation law gives: 
\begin{equation}
(\Omega ,\psi (x_{1},g_{1})\psi (x_{2},g_{2})\Omega )=e^{2\pi is}(\Omega
,\psi (x_{2},g_{2})\psi (x_{1},g_{1})\Omega )  \label{Equ.37}
\end{equation}
for the would-be field-theoretic correlation function which is in agreement
with the expected anyonic statistics of the non-compact localization .

\begin{center}
In order to obtain fields which are localized on semi-infinite strings, one
has to chose a model for F. The choice : 
\begin{equation}
F(g)=e^{isg(0)}  \label{Equ.38}
\end{equation}
with $g=(\gamma ,\omega )$ acting fractionally on the line $u\in R=%
\widetilde{S^{1}}\subset \widetilde{SO(2,1)}$ as: 
\begin{equation}
(\gamma ,\omega )(u)=\omega +\arg \frac{e^{iu}+\gamma }{1+e^{iu}\overline{%
\gamma }}  \label{Equ.39}
\end{equation}
The two-point function specializes to: 
\begin{equation}
\left\langle \psi (x_{1},u_{1})\psi (x_{2},u_{2})\right\rangle =\int \frac{%
d^{2}p}{2\omega }%
e^{ip(x_{1}-x_{2})}e^{-is(L(p)^{-1}(u_{1})-L^{-1}(p)(u_{2}))}  \label{Equ.40}
\end{equation}
and the difference between the left and right hand side in (37 ) replaces
the bosonic commutator function for our anyonic case: 
\begin{equation}
\Delta (\xi ,u_{1},u_{2})=
\end{equation}
\begin{equation}
=\int \frac{d^{2}p}{2\omega }\left\{ e^{ip\xi
}e^{-is(L(p)^{-1}(u_{1})-L(p)^{-1}(u_{2}))}-e^{-ip\xi
}e^{-is(L(p)^{-1}(u_{2})-L(p)^{-1}(u_{1}))}e^{4\pi is(\left[ \frac{%
u_{1}-u_{2}}{2\pi }\right] +\frac{1}{2})}\right\}
\end{equation}
where $\xi $ is the difference of the x's and the square bracket indicates
the nearest larger integer.
\end{center}

$\Delta $ has the property of L-covariance: 
\begin{equation}
\Delta (g\xi ,gu_{1},gu_{2})=\Delta (\xi ,u_{1},u_{2})  \label{Equ.42}
\end{equation}
The vectors $u_{1}$ and $u_{2}$ on the unit circle correspond to a wedge W
and its spacelike complement W'. The simultaneous stability group of u$_{i}$
leaves this wedges invariant. For each $\xi \in W\cup W^{\prime }$ there
exists a transformation K$\in \widetilde{SO(2,1)}$ which is in the conjugacy
class of the $\pi -$rotation which reflects $\xi $ and flips the $%
u_{i}^{\prime }s:$ 
\begin{equation}
K\xi =-\xi \qquad Ku_{1}=u_{2}\qquad K^{2}=2\pi -rotation  \label{Equ.43}
\end{equation}
Under the action of this ''square root'' of the 2$\pi $-rotation the $\Delta 
$ behaves as: 
\begin{equation}
\Delta (K\xi ,Ku_{1},Ku_{2})=\Delta (-\xi ,u_{2},u_{1}+2\pi )=e^{-2\pi
is}\Delta (\xi ,u_{1},u_{2})  \label{Equ.44}
\end{equation}
Consistency between the two transformation formulas yields: 
\begin{equation}
\Delta (\xi ,u_{1},u_{2})=0,\qquad for~~\xi \in W\cup W^{\prime }
\label{Equ.45}
\end{equation}
i.e. as long as u$_{1}$is different from u$_{2}$ any separation $\xi $ of
the string starting points x and y, such that a string crossing is avoided,
will lead to a vanishing $\Delta $ function. A representation in terms of
known functions is presumably easier than for the analog problem of the
d=3+1 continuous spin representation (presented below).

This situation corresponds to light-like strings . The description does not
reveal in a manifest way that these anyons permit the sharper spacelike cone
localization. Only the formation of {\it wave packets } in the {\it %
light-like string direction e} or a covariantization based directly on
space-like strings (using e.g. a de Sitter space representation of the $%
\widetilde{SO(2,1)}$) could reveal sharper localizations inside wedges. We
presented these rather explicit calculations because the covariantization of
the d=3+1 m=0 continuous spin representations in section 6 is completely
analogous, albeit analytically more complicated.

Our argument in favour of spacelike cone localization for anyons is based on
the observation of non-triviality of intersections\cite{Wies} of wedge
spaces H$_{W}^{R}$. Let W$_{1}$and W$_{2}$ be two wedges and form the dense
set of states obtained by averaging with smooth functions of compactly
localized Fourier transform: 
\begin{equation}
\int dsdtf(t,s)\delta _{W_{1}}^{it}\delta _{W_{2}}^{is}\Phi
\,,\,\,\,\,\,\,\,\Phi \in {\cal H}  \label{equ.46}
\end{equation}
This set of vectors is certainly in the domain of $\delta _{W_{1}}^{\frac{1}{%
2}}.$ In order to see that also $\delta _{W_{2}}^{\frac{1}{2}}$can be
applied, we need a commutation relation of $\delta _{W_{2}}^{\frac{1}{2}}$
with $\delta _{W_{1}}^{it}.$ For orthogonal wedges such a commutation
relation is well-known from SO(2,1) group theory: 
\begin{equation}
\delta _{W_{2}}^{\frac{1}{2}}\delta _{W_{1}}^{it}\delta _{W_{2}}^{-\frac{1}{2%
}}=\delta _{W_{1}}^{-it}  \label{equ.47}
\end{equation}
and the case of W's in a more general position may be reduced to this
orthogonal situation and in this way one proves the existence of a
simultaneous dense domain for the $\delta ^{\prime }s$ associated to
different L-boosts

At this point one may be tempted to think that our one-particle analysis,
which relates to the field theoretic two-point function, may be generalized
to the standard commutation relation between creation and annihilation
operators as:

\begin{equation}
a(p,u)a^{*}(q,v)=e^{4\pi is(\left[ \frac{u-v}{2\theta }\right] +\frac{1}{2}%
)}a^{*}(q,v)a(p,u)+2\omega \delta (p-q)e^{-is(L^{-1}(p)(u)-L^{-1}(p)(v))}
\label{equ.48}
\end{equation}
This is however inconsistent (except for bosonic and fermionic phases)
because it can be shown to lead to a contradiction with the associativity of
multiplication for three space-like cone localized anyon operators \cite
{Mund}. The correct multiparticle space from scattering theory has a
different structure from a tensor product Fock space. this was to be
expected fom the nonlinear spin fusion as mentioned before. The anyonic
momentum space creation and annihilation operators associated with
scattering states have source and range projections which have to match the
superselection charges of the state vectors. Lacking a functor from the
Wigner space to von Neumann algebras, one is forced to study the problem of
modular localization of free anyons in the space of incoming scattering
states with conserved number of incoming particles. In such a situation one
expects a ''kinematical'' S-matrix which consists of piecewise constant
phase factors (abelian R-matrices) and hence trivial cross sections\cite
{Schro}. In analogy to factorizing theories in d=1+1 one expects a situation
real particle (on shell) conservation and virtual particle (off shell)
creation. In particular the two point funtion for localizable free anyonic
fields are expected to have continuous contribution beyond the one particle
Wigner contribution. We hope to come back to this interesting problem in the
near future.

As a consequence of the appearance of the directional degrees of freedom for
elementary strings (i.e. strings which cannot be represented in terms of
line integrals of other fields as in Mandelstam type exponential line
integrals) the Wigner anyonic states are more analogous to infinite
component L-covariant wave functions. Their relation to ''Chern-Simons
anyons'' is presently not clear. Note that the geometric Chern-Simons
pictures about anyons suffers from a lack of concreteness: there are no
operator formulas and it is even not clear how to arrive at them. Algebraic
QFT with its intrinsic concepts is expected to be a more suitable place for
their understanding than either via Chern-Simons or through quantum
mechanics.\cite{sch}

Before closing, a brief comment about the d=1+1 situation is in order. In
this case the localization properties of free fields depend on the
''Lorentz-spin'' i.e. on the value of s in the one-dimensional
L-representation factor $\exp s\chi $ with $\chi $ being the rapidity. All s$%
>0$ representations may be obtained in the s=$\frac{1}{2}$ Fock space of
fermions by using appropriate intertwiners u and v. But only for
(half)integer s does one obtain pointlike localized covariant fields. At
generic values one does not get beyond wedge localization. The bad
localization property does not improve in the zero mass limit. The
localizable fields of chiral conformal field theory have a different origin
which is further removed from the Wigner representation theory. They owe
their existence to the peculiar structure of current operators which lead to
Weyl algebras with a nontrivial center. It seems that also the structurally
rich plektonic theories (nonabelian braid group statistics) can also be
traced back to this property \cite{sch}.

\section{ Haag Duality and E.M.Duality.}

Massive free fields obey Haag duality not only for double cones, but also
for topologically more complicated localizations e.g. toroidal regions.
Algebras associated to massless fields for helicity $s\geq 1$ however cause
a topological obstruction resulting in a breakdown of toroidal Haag duality 
\cite{Leyland}. Let {\cal T\ }be the causal completion of a spatial torus we
refer to this Minkowski space region as a ''corona'' \cite{sch} . The size
of the corona is chosen in such a way that the causal complement ${\cal T}$
of consists of a double cone ${\cal T}^{\prime }$ of diameter r and a
''double cone at infinity'': $\left| \vec{x}\right| \geq R+\left| t\right| $
causally separated from the former by the $T$ with width: $R-r\geq 0$ region
in between . Then one obtains the following proper corona- inclusion: 
\begin{equation}
H^{R}{\cal (T)\subset }H^{R}{\cal (T^{\prime })^{\prime }}  \label{Equ.49}
\end{equation}
\[
\curvearrowright {\cal A}({\cal T})\subset {\cal A}({\cal T}^{\prime
})^{\prime } 
\]
where $H^{R}(\cdot )^{\prime }$ denotes the previously defined symplectic
complement and the {\cal A}'s denote the corresponding von Neumann-algebras
as obtained from the $H^{R}(\cdot )$ by the Weyl construction.

This ''classical'' obstruction, formally related to the appearance of $%
\delta ^{\prime }$ in the E-H canonical commutation relation, can be
physically understood in terms of a (suitably regularized) magnetic flux
through a surface which stretches from a circle inside the torus into the
space-like separated region inside. Such a flux does not change if one
passes through another surface subtended from the same circle. Hence such a
flux, also not being localizable within the 4-dim toroidal region
nevertheless belongs to the symplectic complement of the spacelike
complement of the corona consisting of two spacelike seperated pieces. This
entails the above violation of Haag duality for the corresponding algebras.
A more systematic approach in the spirit of Wigner consists in rewriting the
inner product in terms of tensorial object . This time, unlike the massive
case, there are no covariant intertwiners which lead to a nondegenerate
inner product. The best one can do is to introduce a partially covariant
inner product associated (by polarization): 
\begin{eqnarray}
\int \left| \psi (k,\pm )\right| ^{2}\frac{d^{3}k}{2\omega } &=&\int \bar{A}%
_{\mu }(k,n,\pm )A^{\mu }(k,n,\pm )\frac{d^{3}k}{2\omega },\quad \omega
=\left| \vec{k}\right|  \label{Hil} \\
A_{\mu }(k,n,\pm ) &=&\frac{n^{\nu }F_{\nu \mu }(k,\pm )}{n\cdot
k-i\varepsilon },\quad F_{\nu \mu }(k,\pm )=Ak_{\nu }\varepsilon _{\mu
}(k,\pm )\psi (k,\pm )\quad  \nonumber
\end{eqnarray}
where $A$ denotes the antisymmetrization in $\mu ,\nu $ and $\varepsilon
_{\mu }(k,\pm )$ the polarization vectors. The singularity in k-space
corresponds to the semiinfinite line integral along n in x-space. 
\begin{eqnarray*}
A_{\mu }(x,n) &=&\int_{0}^{\infty }n^{\nu }F_{v\mu }(x+ns)ds \\
&=&\int (e^{-ikx}\sum_{i=\pm }A_{\mu }(k,n,i)+h.c.)\frac{d^{3}k}{2\omega }
\end{eqnarray*}
This vector potential has the following obvious properties: 
\begin{eqnarray*}
\partial _{\mu }A_{\nu }-\partial _{\nu }A_{\mu } &=&F_{\mu \nu } \\
(U(\Lambda )A)_{\mu }(x,n) &=&\Lambda _{\mu }^{\nu }A_{\nu }(\Lambda
^{-1}x,n^{\prime }) \\
&=&\Lambda _{\mu }^{\nu }A_{\nu }(\Lambda ^{-1}x,n)+\partial _{\mu }G(x) \\
G(x) &=&\int e^{ikx}\frac{1}{(kn-i\varepsilon )(kn^{\prime }-i\varepsilon )}%
n\cdot F(\Lambda ^{-1}k)\cdot n^{\prime }\frac{d^{3}k}{2\omega }
\end{eqnarray*}
i.e. the Lorentz transformation which acts on the Wigner wave function resp.
on the $F_{\mu \nu }$ tensor, transforms the potential covariantly except an
additive gauge term. The nonlocality of the vectorpotential is made manifest
by this noncovariant transformation law. This peculiar ''gauge'' behaviour
is a consequence of the nonfaithful helicity representation of the
noncompact ''little group'' $E(2).$ In particular{\it \ the quantum origin
of gauge and gauge invariance has nothing to do with the notion of classical
fibre bundles} as most of the books allege. This is one of the more
interesting clashes between quantization and an intrinsic quantum based
approach. The quantization method (from the viewpoint of the Wigner method)
would trade the physical nonlocality of $A_{\mu }$ with (physically
artificial) formal elegance by the introduction of an indefinite metric
(Gupta-Bleuler). In this way the additive term would loose its significance
related to Lorentz transformations and become the gauge concept of the
mathematicians and of classical Maxwell theory. This is the method in which
covariant renormalized perturbation theory is carried out. One profits from
the formal elegance at the prize of a conceptionally questionable return to
quantum physics.

The obstruction against equality in \ref{Equ.49}contains a very interesting
conceptional message. Whereas violations of Haag duality for simply
connected regions are the hallmark of spontaneous symmetry breaking ( in
fact they may be used for a model independent definition of that concept in
the setting of algebraic QFT.), the violation for not (simply) connected
regions has two different physical explanations. The most common one is the
mechanism of ''charge split'' into causally disjoint regions. In this case
the commutant is bigger than the geometric complement suggests because the
charge split mechanism on a neutral observable algebra is not incorporable
into a geometric picture. Of course one is always invited to enlarge the
observable algebra to the field algebra for which there is harmony with the
geometrical picture. The second mechanism is the one at hand: a
''quantum-topology'' caused mismatch between the geometrical complement and
the quantum theoretical opposite in the sense of local commutativity resp.
of symplectic structure. The defect dimension of the two real Hilbert spaces
in \ref{Equ.49}is: 
\[
dim\left[ H^{R}({\cal T}^{\prime })^{\prime }:H^{R}({\cal T})\right] =1 
\]
The obstruction is caused by the presence of just one object: 
\begin{eqnarray*}
&&\oint_{C\subset {\cal T}}A_{\mu }^{reg}(x,n)dx^{\mu } \\
A_{\mu }^{reg}(x,n) &=&\int \rho (\vec{x}-\vec{y})A_{\mu }(\vec{y}%
,x_{0},n)d^{3}y
\end{eqnarray*}
The integration is over a closed path $C$ inside ${\cal T}$ and we
regularized the vector potential with a smooth function of small support $%
supp\rho \in B_{\varepsilon }$ so that one maintains normalizability \ref
{Hil}and remains inside ${\cal T}.$ The line integral represents the class
of expressions of this kind, any two such elements differ only by field
strength localized in ${\cal T}.$ The line integral is a L-invariant and may
be expressed in terms of a magnetic flux through any surface $S$ with the $C$
boundary. It is precisely this floating surface stretching beyond $C,$ which
in the quantum setting of commutativity (or symplectic orthogonality)
prevents the affiliation with $H^{R}(T)$ and makes it a member of the
nongeometric $H^{R}(T^{\prime })^{\prime }.$ This is of course an intrinsic
property of the theory which cannot be removed by the indefinite metric
formalism.

It is a much more difficult question as to what becomes of this topological
obstruction in the presence of interactions. The reader can find some
remarks in the last section. It is tempting to interpret this obstruction as
indicating the necessity of an interaction \cite{Haag} i.e.of the presence
of non-vanishing electric or magnetic (or both) currents. 
\begin{equation}
\partial ^{\mu }F_{\mu \nu }(x)=j_{\nu }(x),\,\,\,\partial ^{\mu }\tilde{F}%
_{\mu \nu }(x)=\tilde{j}_{\nu }(x)  \label{Equ.50}
\end{equation}
The idea is that interactions are necessary to restore perfect Haag duality
which is violated in the free theory. Such a point of view would attribute a
very distinguished role to electromagnetic duality i.e. those superselection
rules which originate from the quantum version of the Maxwell structure and
may well be the physical concept behind the semi-classical ''gauge
principle''.This issue of problematizing the notion of ''magnetic field'' on
the same level of depth as the notion of ''charge'' in the DHR
superselection theory is presently ill-understood in QFT.

In low dimensional QFT the analogous issue of order-disorder duality and the
connection with Haag duality is much better understood. There, even in free
theories, it is not possible to have {\it \ no} charge sectors with both
order and disorder the realization of both charges being related in d=1+1 to
the zero mass limit. The previous idea of maintaining corona duality would
bring the interacting Maxwell-like theories closer to the 2-dim.
situation.This analogy is another reason to believe that the free Maxwell
situation is peculiar. The remaining three non-peculiar cases, namely the
appearance of objects with e.-, m.- or e.m.-charges have an infrared
structure, whose implications for localization properties are outside the
present scope of understanding. A better understanding of the connection
between these properties and the modular theory (in the vein of the remarks
about interactions in the last section) seems to be essential for future
progress..

The corona inclusion may be constructed solely in terms of the Wigner theory
supplemented by the modular theory for wedges avoiding covariant amplitudes
like $F_{\mu \nu }$ alltogether. Helicities h$\geq 1$ present similar corona
structures.

\section{ The Localization Properties of the Positive Energy Continuous Spin
Representations.}

Already in the late 60's the question of how to covariantize and incorporate
Wigner's zero mass continuous spin representation into existing frameworks
of QFT arose some interest notably with physicists who were familiar with
the spirit of algebraic QFT \cite{Mack}. Whereas in standard QFT based on
quantization and Lagrangians these representations were usually dismissed as
uninteresting because ''nature appearently does not make use of them'', the
spirit in which algebraic field theorists approached this problem was more
''Wignerian''. They asked whether these representations fulfill the
localization properties which are inexorable attributes (in addition to
their indecomposability expressed in the irreducibility requirement)of
particles. It was found that they do not permit a compact localization as
the standard Wightman fields do.

Looking again at the old computations with the hindsight of the non-compact
space-like cone localization of section 4, one easily realizes that their
natural covariantization leads to the same wedge-localized light-like
strings with an infinite component unitary representation on light-like
directions and complex variables $\xi _1,\xi _2$ taking over the role of the
u in the case of the anyonic representation of section 4. The identification
of complex 2-vectors with light-like directions is done with the help of the
Pauli matrices: 
\begin{equation}  \label{Equ.51}
l_\mu =\xi ^{+}\sigma _\mu \xi
\end{equation}
the intertwiners for this light-like covariantization are: 
\begin{equation}  \label{Equ.52}
u(p,\xi )=f_\lambda ^{\rho \chi }(\xi B_p),\,\,\,\,\,\,\,f_\lambda ^{\rho
\chi }(\xi )=\left| \xi _2\right| ^{2c-2}e^{-i\lambda \Phi _1}J_{l_0-\lambda
}(\frac \rho \kappa \left| z\right| )e^{il_0\Phi _2}
\end{equation}
\[
z=\frac{\xi _1}{\xi _2},\,\,\,\,\Phi _{1,2}=\pm \pi +\arg \xi _1\mp \arg \xi
_2 
\]
The $f_\lambda ^{\rho \chi }(\xi )\,$ corresponds precisely to the
equivariant function F in section\thinspace 4.

The step from wedge-localization to space-like cone localization is also
analogous to section 4. Instead of two orthogonal wedges one now considers
three. The smoothening with testfunctions of compact Fourier-transform: 
\begin{equation}  \label{Equ.53}
\int dtdsduf(t,s,u)\Delta _1^{it}\Delta _2^{is}\Delta _3^{iu}\Phi ,\,\,\Phi
\in {\cal H}
\end{equation}
together with commutation relations between the three boosts analogous to
the ones used in section 4 will give a simultaneous dense domain for $\Delta
_i^{\frac 1{2.}}$ i=1,2,3\thinspace. However it is not clear if the case of
three wedges in general position can be reduced to the orthogonal case.

It is interesting to compare this space-like cone localization with that
established by Buchholz and Fredenhagen on the basis of the spectral gap
assumption\cite{Haag}. In their massive case it is not possible to realize
this in a theory with a on mass-shell supported two-point function.

\section{\thinspace Intrinsic Understanding of Interactions? Programmatic
Remarks.}

Since the wedge regions have a preferential status with respect to the
construction of interaction-free algebras, it is tempting to think that this
may be helpful in obtaining some intrinsic insight into interactions. Let us
take a helping hand from scattering theory. There it is shown that out- and
ingoing- fields share the same Poincar\'e transformations with the
interacting fields i.e. they both possess the same modular transformations $%
\Delta ^{it}$ for the wedge region. Only the TCP conjugation is sensitive
with respect to interactions. In order to see this we will derive the
following representation for S which is valid in an asymptotically complete
theory : 
\begin{equation}  \label{Equ.54}
S=\theta \cdot \theta _0=J\cdot J_0
\end{equation}

$J$= modular conjugation for interacting wedge algebra

$J_{0}$ = modular conjugation for the interaction- free incoming wedge
algebras.

The formula follows from the $\Theta =$TCP transformation of Heisenberg
fields. Taking the LSZ limit on this transformation formula and noticing
that both sides approach different (in and out) limits ( and remembering
that the spatial rotation factors between $\Theta $ and J are independent of
interactions), one obtains the above representation.

The interacting and incoming wedge algebras are of the same type, in fact
they are expected to be type III$_{1}$factors\ .Since both are living in the
same Hilbertspace, their isomorphism amounts to a unitary equivalence.
Unfortunately this kind of argument does not lead to a ''natural'' unitary
operator which we expect to be some kind of natural ''square root'' of S
i.e. some kind of ''algebraic'' M\"{o}ller operator. On a very formal level
the method of Bogoliubov and Shirkov leads to such unitaries, but this would
bring us back to the interaction picture and the formal time ordered
operator expressions $S(g)$.

Let us therefore be more modest and just ask for a modular net of real
Hilbert-subspaces of the incoming Fock-space. If we pose this problem in two
space-time dimensions, we could take a $J$ operator which is different from $%
J_{0}$ by one of those rather simple rapidity dependent factorizing
S-matrices of the ''bootstrap construction'' which are the long-distant
limits of the class of theories with the same superselection rules \cite{sch}%
. Here our modular proposal is expected to give a more field theoretic
understanding of the so-called formfactor bootstrap program and the
Bethe-ansatz approach. Both the formfactor program and the present ''modular
program'' point into the same direction: the construction of local fields
resp. of local nets from a given S-matrix. Whereas the formfactor program
has only been formulated for factorizable S-matrices, the modular idea in
principle does not suffer from such a restriction. Presently it is not known
if the latter leads to a unique net of {\it wedge algebras}; the above
argument only yields unique net of real local {\it wedge subspaces} of the
Fock space of scattering states. The uniqueness modulo normalizations
(corresponding to the different composite fields) of the formfactor program
suggests strongly the uniqueness of the general inverse scattering problem
in algebraic QFT.

Note that in higher dimensions such a starting point with a model S-matrix
is not available since the above long distance limits give S=1 and hence
free field equations (it may however lead to a solid proof of the underlying
''folklore'' statement that S=1 leads necessarily to the free field Borchers
class.) in agreement with the d=3+1 Coleman-Mandula theorem or the theorem
that interaction in the sense of $S\neq 1$always implies the presence of
inelastic real processes.

In this case one could contemplate to start with a unitary
Poincar\'{e}-invariant operator $S_{aux}$\ which only fulfills the
TCP-invariance and the cluster decomposition property. This alone already
leads to the existence of a net of wedge subspaces of the Fockspace: {\cal H}%
$_{W,F}^{R}\subset {\cal H}_{F}\,$. The requirement that these wedge spaces
contain real subspaces describing states localized in noncompact space-like
cones or compact double cones i.e. that certain intersections of wedge
spaces are nonempty, is expected to yield analytic on shell restrictions on $%
S$ that require corrections on $S_{aux}$. One would hope that such
restrictions resulting from localizations which are sharper than the
original wedge localization may give rise to an inductive procedure (a kind
of algebraic perturbation theory in which the unitarity relations are
fulfilled in every order). Even if such a program succeeds to yield a net of
local subspaces of ${\cal H}_{F}$, it is not clear that a functor from real
subspaces of the Fock space exists (thus generalizing the free CCR and CAR
functors). Nevertheless since one is combining the deep modular properties
of algebraic QFT with the startling nonperturbative successes of low
dimensional QFT this way of thinking offers an irresistable temptation. The
starting auxiliary $S_{aux}$\ could be a unitary operator as in Heisenberg's
ill-fated S-matrix theory \cite{Heisenberg} but unlike in Heisenberg's
attempt or in Chew's bootstrap proposal the $S-$matrix would be totally
subservient to QFT.

The use of S-matrix properties for the purpose of constructing localized
fields and algebraic nets is however not available if infrared properties
wreck the mass shell i.e. if the LSZ-asymptotes vanish as in the case of
particles carrying Maxwellian charges. Whereas these infrared problems are
no serious obstacle for theories which can be considered as scaling limits
of massive theories with spectral gaps ( e.g. 2-dim. conformal theories as
limits of integrable massive models), Maxwellian theories as in section 5
pose a serious problem since the lack of a particle reference space and a
standard $S-$matrix r magnetic) infrared problems prevent the
straightforward use of modular localization. In fact the structural insight
into such theories is so poor that even nonperturbative arguments in favour
of the relation of spin and statistics and TCP are unknown.

A particularily simple distinguished class of S-matrices for this program
are the piecewise constant matrices which were mentioned in section 4. In
d=1+1 this is related to the construction of (dis-)order fields associated
to the free field Borchers class. In d=2+1 however ''free anyons'' cannot be
represented in the Fock-space of Fermions and Bosons by such simple
modifications of the free Borchers class and one expects their formfactor
construction to require concepts similar to the factorizing models even if
their S-matrix (believed to be piecewise constant) is simpler \ref{Schro}.

Very recently some significant progress on this problem was obtained through
the discovery of the ''microlocal spectrum condition'' which also permits to
take into account interactions between the matter fields\cite{Brunetti}.
Although the perturbation theory is not based on quantization (it rather
uses the Bogoliubov Weinberg dispersion theoretic framework refined by the
Epstein-Glaser theory), it is not as intrinsic as a modular-based approach
proposed (but unfortunately not carried out) here. The time-ordering used in
that formalism originates from Dirac's formalism for time dependent
Hamiltonian perturbations. Although being a natural part of quantum
mechanics and hence independent of (canonical, functional)quantization ,
such concepts using the 4$^{th}$component of a vector and distinguishing
hyperplanes are not a good starting point for an {\it intrinsic} approach to
interactions. The modular wedge theory however is a necessary consequence of
covariant nets. It is not only characteristic for the latter, but also
explaines and underlines the big distance which QFT maintains to classical
field theory as well as to quantum mechanics.

The modular approach to interactions advocated here may also be useful for a
more intrinsic understanding of renormalization. The usual presentation of
renormalization is intimately related to quantization and actions. It only
repairs a very formal and slightly illegitimate starting point\cite{S}.
Looking only at the renormalized correlation functions, it is not so easy to
isolate an intrinsic aspect of renormalization \cite{S}. Even in Wilson's
renormalization group approach the intrinsic characterization of fix points
outside of Gaussians remains unclear.

In the problem of d=2 critical indices an intrinsic understanding has
finally be achieved in terms of very subtle properties of their attached
noncommutative real time chiral theories\cite{S}. It turned out that the
critical indices are classified by certain numerical values of superselected
charges (related to the so called ''statistical dimensions'' of algebraic
QFT) which are in turn related to such (at first sight) remote looking
issues as the classification of physically admissible braid-group statistics
and modular theory. An intuitive understanding in terms of values of charges
appears first in Kadanoff''s work\cite{Kad}.

Recently a framework was proposed which allows to understand an intrinsic
association of a scale invariant theory to a massive theory (with the
possibility of new superselection rules emerging in the short distance limit
i.e. short-distance quark deconfinement)\cite{Bu}. I believe that this
framework together with ideas from modular theory may also cast some light
on a possible distinction between ''renormalizable and unrenormalizable
nets''. In addition there is the interesting issue of ''semirenormalizable''
theories i.e. the question of how to deal with theories like massive (non
Higgs) vectormesons coupled to charged matter fields via conserved currents.
In that case the neutral fields stay renormalizable. To have (observable)
renormalizable subsets of fields could very well (as the causality issue) be
a general phenomenon of higher spin interactions.

Finally we want to emphasize that our modular proposal based on the S-matrix
does not cover zero mass theories. Whereas for those massless theories which
can be viewed as scaling limits of massive theories ( e.g. chiral conformal
QFT ) this poses no serious problem (since the conceptual complication is
compensated for by an analytic simplification), the physically interesting
cases in which the dual e.and m. charges are of Maxwellian origin remain
presently outside the modular approach. In this case neither the conceptual
nor the analytic aspects are simple.

If in these notes I created the impression that QFT, despite its more than
60 years of existence, is a very young and fresh branch of intellectual
endavour (looking at the many basic but insufficiently understood problems),
then this was not without intentions.

A large part of this work ( the main exception being the section on
''continuous spin'') was carried out during a visit of the UFES Brazil. I am
deeply indebted to some of my colleagues at the UFES, in particular to
proof. Julio Cesar Fabris for their kind hospitality. I am also indebted to
detlev Buchholz for several valuable informations.

\end{document}